\documentclass[twocolumn,showpacs,preprintnumbers,amsmath,amssymb]{revtex4}
\usepackage{graphicx}
\usepackage{dcolumn}
\usepackage{bm}
\begin{document}
%
\def \chic#1{{\scriptscriptstyle #1}}
\def \med#1{{\scriptstyle #1}}
\def \chicl{{\chic L}}
\def \ii{\'{\i}}
\def \vsk{\vspace{0.52cm}}
\def \vseq{\vspace{0.05cm}}
\def \be{\begin{equation}}
\def \ee{\end{equation}}
\def \bea{\begin{eqnarray}}
\def \eea{\end{eqnarray}}
\def \ba{ \begin{array}}
\def \ea{ \end{array}}
\def \ni{\noindent}
\def \lb{\label}
\def \En{E_\nu}
\def \dEn{{\med{(E_\nu)}}}
\def \sigask{ \sigma^a_{\chic{SK}}\dEn }
\def \sigesk{ \sigma^e_{\chic{SK}}\dEn }
\def \sigeask{ \sigma^{e,a}_{\chic{SK}}\dEn }
\def \sigessno{ \sigma^{\chic{ES}}_{\chic{SNO}}\dEn }
\def \sigccsno{ \sigma^{\chic{CC}}_{\chic{SNO}}\dEn }
\def \signcsno{ \sigma^{\chic{NC}}_{\chic{SNO}}\dEn }
\def \sigesno{ \sigma^e_{\chic{SNO}}\dEn }
\def \sigeasno{ \sigma^{e,a}_{\chic{SNO}}\dEn }
\def \Pecal{{\cal P}_e }
\def \Pacal{{\cal P}_a }
\def \Pscal{{\cal P}_s }
\def \Peacal{{\cal P}_{e,a}}
\def \phie{ \phi^{\nu_e} }
\def \phia{ \phi^{\nu_a} }
\def \nn{\nonumber}
%
%
\title{Active Neutrino Oscillations and
the SNO Neutral Current Measurement}

\author{Alexis A. Aguilar-Arevalo}
 \email{alexis@nuclecu.unam.mx}
\author{J. C. D'Olivo}
 \email{dolivo@nuclecu.unam.mx}
\affiliation{
Instituto de Ciencias Nucleares,
Universidad Nacional Aut\'onoma de M\'exico,
Apartado Postal 70-543, 04510 M\'exico, Distrito Federal, Mexico
}

\date{\today}

\begin{abstract}
We discuss the relation between the observed CC, ES, and NC fluxes with the
flavor fractional content of the solar neutrino flux seen by SNO.
By using existing estimates of the
cross sections for the charged and neutral current reactions which take
into account the detector resolution, we show how the forthcoming SNO rates
unconstrained by the standard $^8\!$B shape could test the oscillations
into active states. We perform a model independent analysis
for the Super-K and SNO data, assuming a non distorted spectrum.
\end{abstract}

\maketitle


\section{Introduction}

Recently, the SNO collaboration has presented the first direct
measurement of the total active flux of $^8\!$B neutrinos comming
from the Sun \cite{sno_nc}. SNO detects solar neutrinos by means
of three different reactions: Charged Current reaction (CC) $\nu_e
+ p \rightarrow n + e^+$, Neutral Current reaction (NC) $\nu_x + d
\rightarrow p+n + \nu_x$, and Elastic Scattering reaction (ES)
$\nu_e + p \rightarrow p + \nu_e$. The CC reaction is sensitive
exclusively to $\nu_e$, while he NC and ES reactions are sensitive
to all flavors, with less sensitivity to $\nu_a$ ($a=\mu,\tau$) in
the case of ES. Using the integrated rates above the threshold of
5 MeV for the three reactions, they have determined both, the
electron and the active non-$\nu_e$ component of the $^8\!$B
neutrino flux. The latter is $5.3 \sigma$ grater than zero,
yielding strong evidence for neutrino flavor transformation. This
result has been obtained under the assumption that the shape of
the $^8\!$B neutrino spectrum is the same as predicted by the
Standard Solar Model (SSM) \cite{ortiz}. The absence of a
significant distortion of the spectrum has been observed by
Super-Kamiokande (SK) \cite{superk} and confirmed by SNO. The
impact of recent SNO results on the global oscillations solutions,
including all solar neutrino data, have been analyzed by several
authors \cite{global,bandy1}.

 In this work we address the question of how the
forthcoming SNO rates unconstrained by the standard $^8\!$B shape
can be used to test the presence of non-electron active neutrinos
in the solar neutrino flux. In Sec. \ref{primera} we establish the
relation between the fractional flavor components of the spectrum
$P_{e,a}$ and the quantities $\Peacal$ that determine the CC, NC,
and ES fluxes in terms of the measured experimental rates. This
relation involves an average over the appropriate experimental
response functions and are presented in Sec. \ref{segunda}. In
this section we also illustrate to what extent the SNO rates
unconstrained by the standard $^8\!$B shape could play a role in
testing the active oscillations hypothesis.
 A model independent
analysis of the data of SNO and SK incorporating the NC
measurement of SNO is given in Sec. \ref{ultima}.

\section{Energy Spectrum at SNO}
\lb{primera}

The count-rate per energy interval at SNO for the NC events is
related to the true (and unknown) spectrum of solar neutrinos arriving
at the Earth $\phi\dEn$, as follows:
\be
\lb{uno}
\frac{dR^{\chic{NC}}_{\chic{SNO}}}{d\En} = \phi\dEn \signcsno
\left[  P_{e}\dEn + P_a\dEn \right] \; ,
\ee

\ni where $\signcsno$ is the cross section for the NC process
 and $P_{e,a}\dEn = \phi^{\nu_{e,a}}\dEn / \phi\dEn$. The quantities
$P_x\dEn$ ($x=e,a,s$) satisfy $\sum_x P_x\dEn=1$, with $s$ denoting
a sterile neutrino. According to the SSM \cite{bp:2001} the only neutrinos
produced in the Sun are $\nu_e$, therefore the neutral current
count rate at SNO should be
\be
\frac{dR^{\chic{NC}}_{\chic{SSM}}}{d\En} = \phi_{\chic{SSM}}\dEn \;
\signcsno \; ,
\ee

\ni where $\phi_{\chic{SSM}}\dEn$ is the energy spectrum of the $\nu_e$
given by the SSM.

Let $f$ be the ratio of the {\it true} total neutrino flux $\phi $ to the
predicted total flux $\phi_{\chic{SSM}}$:
\be
\phi = f\; \phi_{\chic{SSM}} \; .
\ee

\ni We say that there is \it no deformation \normalfont of the
neutrino spectrum \it produced in the Sun \normalfont with respect to
the SSM prediction if $\phi\dEn = f \; \phi_{\chic{SSM}}\dEn$. In a
more general situation, we could have
\be
\lb{deformation_def}
\frac{\phi\dEn}{f\; \phi_{\chic{SSM}}\dEn}= \zeta\dEn \ne 1 \; ,
\ee

\ni with $\zeta\dEn$ a certain positive function of $\En$, that
satisfies $\int d\En \;\zeta\dEn \;\phi_{\chic{SSM}}\dEn = \phi_{\chic{SSM}}$.
We have assumed that only $\nu_e$ are produced in the Sun.
In general, the ratio $r^{\chic{NC}}_{\chic{SNO}}\dEn$ of the observed to
the theoretical neutral current spectra will be energy-dependent:
\be
\lb{ratio_nc}
r^{\chic{NC}}_{\chic{SNO}}\dEn =
\frac{dR^{\chic{NC}}_{\chic{SNO}}/d\En}
{dR^{\chic{NC}}_{\chic{SSM}}/d\En} =
f \left[  \Pecal\dEn +  \Pacal\dEn \right] \; ,
\ee

\ni where
\be
\lb{def_peacal}
{\cal P}_x\dEn = \zeta\dEn \; P_{x}\dEn \; ,
\ee

\ni and $\sum_x{\cal P}_x\dEn=\zeta\dEn$. If the neutrino spectrum
produced in the Sun has no deformation, then the function
$\zeta\dEn$ in (\ref{deformation_def}) is equal to one for all
energies. In this case, ${\cal P}_x\dEn =
 P_{x}\dEn$,
and $\sum_x \cal P\it_x\dEn = \rm 1$.

Let $\varphi_{\chic{SSM}}\dEn=
\phi_{\chic{SSM}}\dEn/\phi_{\chic{SSM}}$ denote the normalized
solar  neutrino spectrum predicted by the SSM. This quantity
satisfies the relation $\varphi_{\chic{SSM}}\dEn {\cal P}_x \dEn =
\varphi\dEn P_{x}\dEn$, where $\varphi\dEn=\phi\dEn/\phi$ is the
true (and unknown) normalized solar neutrino spectrum. The
integrals over the relevant energy range of the normalised spectra
are equal to one
\be
\int d\En\;\varphi\dEn = \int d\En \;\varphi_{\chic{SSM}}\dEn = 1
\;.
\ee
From the fact that $P_e\dEn + P_a\dEn \leq 1$, we have \bea \int
d\En \;\varphi_{\chic{SSM}}\dEn\; \left[\Pecal\dEn +
\Pacal\dEn\right] = \hspace{2cm} \nn  \\ \int d\En\;\varphi \dEn\;
\left[P_e \dEn + P_a \dEn\right] \leq 1 \; . \hspace{0.4cm} \eea
\ni and therefore, if $\Pecal\dEn+\Pacal\dEn$ is a constant, we
have $0\leq\left[\Pecal\dEn + \Pacal\dEn \right]\leq 1$. In
addition, if all the ${\cal P}_x$ are constant then $\sum_x {\cal
P}_x = 1$.

The ratio of the observed to the predicted charged current spectra
can also be written as
\bea
\lb{ratio_cc}
r^{\chic{CC}}_{\chic{SNO}}\dEn &=& \frac{dR^{\chic{CC}}_{\chic{SNO}}/
dE_\nu}{dR^{\chic{CC}}_{\chic{SSM}}/ d\En} \nn \\
&=&
\frac{\phi\dEn \; \sigccsno\; P_e\dEn}{\phi_{\chic{SSM}}\sigccsno} =
f \Pecal\dEn \; ,
\eea

\ni where $\sigccsno$ is the cross-section for the CC
reaction. Relations (\ref{ratio_nc})  and (\ref{ratio_cc}) are model
independent. They make no assumption on $f$ or neutrino oscillations,
nor require the quantities $\Peacal\dEn$ to be considered as
probabilities.

The elastic scattering event rate is also available from SNO.
This rate, normalised to the SSM prediction is given by
\bea
\lb{ratio_es}
r^{\chic{ES}}_{\chic{SNO}}\dEn &=& \frac{dR^{\chic{ES}}_{\chic{SNO}}/
dE_\nu}{dR^{\chic{ES}}_{\chic{SSM}}/ d\En} \nn \\
&=& f \left( \Pecal\dEn + \rho\;\Pacal\dEn \right) \; ,
\eea

\ni where $\rho=\sigma^{a}_{\chic{SNO}}\dEn/
\sigma^{e}_{\chic{SNO}}\dEn\approx 0.154$
for $E_\nu\geq5$ MeV.

Using Eqs. (\ref{deformation_def}) and (\ref{def_peacal}), the
$\nu_e$ component of the solar neutrino flux
$\phie\dEn=P_e\dEn \; \phi\dEn$
can be written as
\be
\lb{phi_f_pcal} \phie\dEn = f \; \Pecal\dEn \;
\phi_{\chic{SSM}}\dEn \; . \ee

\ni We will say that the electron neutrino spectrum has {\it no
deformation at the Earth} whenever $\phie\dEn$ is proportional to
$\phi_{\chic{SSM}}\dEn$. Then, from Eq. (\ref{phi_f_pcal}) we see
that a constant $\Pecal$ would imply that there is no distortion
of the $\nu_e$ spectrum at the Earth, and viceversa.

According to SK \cite{superk} and SNO \cite{sno_cc,sno_nc} the ratios
$r^{\chic{CC}}_{\chic{SNO}}\dEn$, $r^{\chic{NC}}_{\chic{SNO}}\dEn$,
and $r^{\chic{ES}}_{\chic{SNO}}\dEn$ are practically constant for
$E_\nu\geq$ 5 MeV. As a consecuence, $\Peacal\dEn$  are constants
as can be seen by  taking any combination of two equations among
(\ref{ratio_nc}), (\ref{ratio_cc}), and (\ref{ratio_es}).
For example, from Eqs. (\ref{ratio_nc}), and (\ref{ratio_cc}) we have
\be
\lb{sno_constant_ratio}
\Pecal  = \frac{r^{\chic{CC}}_{\chic{SNO}}}{f} \;,\;\;
\Pacal  = \frac{1}{f}\left(r^{\chic{NC}}_{\chic{SNO}}
- r^{\chic{CC}}_{\chic{SNO}} \right) \; ,
\ee

\ni with $r^{\chic{CC}}_{\chic{SNO}}$,
$r^{\chic{NC}}_{\chic{SNO}}$, and $\Peacal$ constants. Therefore,
the present experimental evidence indicates that no significant
distortion of the $^8\!$B neutrino spectrum has been observed at
the Earth. In principle, in Eq. (\ref{phi_f_pcal}) the energy
dependence of the {\it true} survival probability $P_e\dEn$ could
be approximately compensated by $\zeta\dEn$ in order to explain
the observed energy independence of the neutrino spectrum at the
Earth. Therefore, a distortion of the neutrino spectrum produced
in the Sun remains as an unlikely speculation.

\section{SNO fluxes}
\lb{segunda}

 The elastic scatering rate
measured by SNO can be written in the form
\be
\lb{gg}
R^{\chic{ES}}_{\chic{SNO}} = \overline\sigma^{\;\chic{ES}}
_{\chic{SNO}}\;
\phi^{\chic{ES}}_{\chic{SNO}} \; , \;\;
\ee
\ni with
\be
\lb{prob_av_es}
\ba{rll}
\phi^{\chic{ES}}_{\chic{SNO}}&=& \phi \;
\left[ \langle \Pecal\rangle^{\chic{ES}}_{\chic{SNO}}
+ \rho\;\langle \Pacal\rangle^{\chic{ES}}_{\chic{SNO}} \right] \; ,
\vspace{0.1cm} \\
\overline\sigma^{\;\chic{ES}}_{\chic{SNO}} &=& \int dE_\nu \:
\varphi_{\chic{SSM}}\dEn \;
\sigma^{\;\chic{ES}}_{\chic{SNO}}\dEn \; ,
\vspace{0.1cm} \\
\langle \Peacal\rangle^{\chic{ES}}_{\chic{SNO}} &=&
\frac{1}{\overline\sigma^{\;\chic{ES}}_{\chic{SNO}}}
\int dE_\nu \;\varphi_{\chic{SSM}}\dEn \;
\sigma^{\;\chic{ES}}_{\chic{SNO}}\dEn\;\cal P\it_{e,a}\dEn\;.
\ea
\ee

\ni Here, $\phi^{\chic{ES}}_{\chic{SNO}}$ is the measured elastic
scattering flux.

With similar definitions, the CC event count-rate
is given by
\be
\lb{ff}
R^{\chic{CC}}_{\chic{SNO}} = \overline\sigma^{\;\chic{CC}}
_{\chic{SNO}}\;
\phi^{\chic{CC}}_{\chic{SNO}} \; ,
\ee
\ni where
\be
\lb{prob_av_cc}
\ba{rll}
\phi^{\chic{CC}}_{\chic{SNO}}&=& \phi \;
\langle \Pecal\rangle^{\chic{CC}}_{\chic{SNO}} \; ,
\vspace{0.1cm} \\
\overline\sigma^{\;\chic{CC}}_{\chic{SNO}} &=& \int dE_\nu \:
\varphi_{\chic{SSM}}\dEn \;
\sigma^{\;\chic{CC}}_{\chic{SNO}}\dEn \; ,
\vspace{0.1cm} \\
\langle \Pecal\rangle^{\chic{CC}}_{\chic{SNO}} &=&
\frac{1}{\overline\sigma^{\;\chic{CC}}_{\chic{SNO}}}
\int dE_\nu \;\varphi_{\chic{SSM}}\dEn \;
\sigma^{\;\chic{CC}}_{\chic{SNO}}\dEn\;\cal P\it_e\dEn\;.
\ea
\ee

\ni In Eq. (\ref{ff}), $\phi^{\chic{CC}}_{\chic{SNO}}$ is the  flux
measured by SNO through the CC reaction.

The electron neutrino component of the flux seen by SNO through the elastic
scattering reaction is
\be
\lb{esta}
(\phi^{\chic{ES}}_{\chic{SNO}})_{\chic{\nu_e}}
= \phi\;\langle \Pecal \rangle^{\chic{ES}}_{\chic{SNO}}\; .
\ee

\ni From (\ref{prob_av_cc}) and (\ref{esta}) we get
\be
\lb{aa}
\frac{\phi^{\chic{CC}}_{\chic{SNO}}}{\phi^{\chic{ES}}_{\chic{SNO}}} =
\frac{(\phi^{\chic{ES}}_{\chic{SNO}})_{\chic{\nu_e}}}{\phi^{\chic{ES}}
_{\chic{SNO}}} \times
\frac{\langle \Pecal \rangle^{\chic{CC}}_{\chic{SNO}}}
     {\langle \Pecal \rangle^{\chic{ES}}_{\chic{SNO}}} \; .
\ee


The event count-rate for the NC  can be written as follows:
\be
\lb{ee}
R^{\chic{NC}}_{\chic{SNO}} = \overline\sigma^{\;\chic{NC}}
_{\chic{SNO}}\;
\phi^{\chic{NC}}_{\chic{SNO}} \; ,
\ee
\ni where we have defined
\be
\lb{prob_av_nc}
\ba{rll}
\phi^{\chic{NC}}_{\chic{SNO}}&=& \phi \;
\left[ \langle \Pecal\rangle^{\chic{NC}}_{\chic{SNO}}
                     + \langle \Pacal\rangle^{\chic{NC}}_{\chic{SNO}}
\right] \; ,
\vspace{0.1cm} \\
\overline\sigma^{\;\chic{NC}}_{\chic{SNO}} &=& \int dE_\nu \:
\varphi_{\chic{SSM}}\dEn \;
\sigma^{\;\chic{NC}}_{\chic{SNO}}\dEn \; ,
\vspace{0.1cm} \\
\langle \Peacal\rangle^{\chic{NC}}_{\chic{SNO}} &=&
\frac{1}{\overline\sigma^{\;\chic{NC}}_{\chic{SNO}}}
\int dE_\nu \;\varphi_{\chic{SSM}}\dEn \;
\sigma^{\;\chic{NC}}_{\chic{SNO}}\dEn\;\cal
P\it_{e,a}\dEn\;.
\ea
\ee

\ni Here, $\phi^{\chic{NC}}_{\chic{SNO}}$ represents the flux measured
by SNO through the NC reaction.  We must keep in mind that the cross
sections $\sigma^{\;\chic{ES}}_{\chic{SNO}}\dEn$,
$\sigma^{\;\chic{CC}}_{\chic{SNO}}\dEn$, and
$\sigma^{\;\chic{NC}}_{\chic{SNO}}\dEn$,
that appear in Eqs. (\ref{prob_av_es}), (\ref{prob_av_cc}), and
(\ref{prob_av_nc}) depend on the response
functions of the SNO detector.

If $(\phi^{\chic{NC}}_{\chic{SNO}})_{\chic{\nu_e}}
=\phi\;\langle \Pecal \rangle^{\chic{NC}}_{\chic{SNO}}$ is the electron
neutrino component of the flux seen by SNO through the NC
reaction, then from (\ref{prob_av_cc}) it is clear that
\be
\lb{otra}
\frac{\phi^{\chic{CC}}_{\chic{SNO}}}{\phi^{\chic{NC}}_{\chic{SNO}}} =
\frac{(\phi^{\chic{NC}}_{\chic{SNO}})_{\chic{\nu_e}}}{\phi^{\chic{NC}}
_{\chic{SNO}}} \times
\frac{\langle \Pecal \rangle^{\chic{CC}}_{\chic{SNO}}}
     {\langle \Pecal \rangle^{\chic{NC}}_{\chic{SNO}}} \; .
\ee

A ratio $(\phi^{\chic{ES}}_{\chic{SNO}})_{\chic{\nu_e}}
/\phi^{\chic{ES}}_{\chic{SNO}}$ less than one necessarily implies
the presence of a non-$\nu_e$ active neutrino in the solar
neutrino flux. What can actually be done with the experimental
measurements is to calculate the ratio
$\phi^{\chic{CC}}_{\chic{SNO}}/\phi^{\chic{ES}}_{\chic{SNO}}$. As
Eq. (\ref{aa}) shows, in principle it could be possible to have
the ratio
$(\phi^{\chic{ES}}_{\chic{SNO}})_{\chic{\nu_e}}/\phi^{\chic{ES}}
_{\chic{SNO}}$ equal to one, and still be in agreement with the
experimental results from SNO by having $\langle \Pecal
\rangle^{\chic{CC}}_{\chic{SNO}}/ \langle \Pecal
\rangle^{\chic{ES}}_{\chic{SNO}} < \rm 1$. However, given the
observed non-dependency of the quantities $\Pecal\dEn$ on the
energy, we have that the averages defined in Eqs.
(\ref{prob_av_es}) and (\ref{prob_av_cc}) are approximately equal:
$\langle \Pecal\rangle^{\chic{CC}}_{\chic{SNO}} \approx \langle
\Pecal\rangle^{\chic{ES}}_{\chic{SNO}} \approx\Pecal$. When this
result is combined with Eq. (\ref{aa}), gives irrefutable evidence
that there are $\nu_\mu$ and/or $\nu_\tau$ arriving at the
detector. A similar conclusion can be drawn by comparing the CC
and NC fluxes. The experimental evidence suggests that
$\langle\Pecal\rangle^{\chic{CC}}_{\chic{SNO}}\approx
\langle\Pecal\rangle^{\chic{NC}}_{\chic{SNO}}\approx\Pecal$, from
where we see that
$\phi^{\chic{CC}}_{\chic{SNO}}/\phi^{\chic{NC}}_{\chic{SNO}}<1$
implies
$(\phi^{\chic{NC}}_{\chic{SNO}})_{\chic{\nu_e}}/\phi^{\chic{NC}}
_{\chic{SNO}} < 1$.

The $CC/NC$ ratio of rates given by the SNO collaboration has been derived
assuming the SSM $^8\!$B spectral shape. Up to now SNO has not released the
information for the corresponding unconstrained ratios. When this information
becomes available the absence of active neutrino flavor transformations
could be ruled out even for a non constant $\Pecal\dEn$. To see this,
let us assume for a moment that $P_a\dEn=0$. Then, we have
\be
\frac{(\phi^{\chic{ES}}_{\chic{SNO}})_{\chic{\nu_e}}}{\phi^{\chic{ES}}
_{\chic{SNO}}} =
\frac{(\phi^{\chic{NC}}_{\chic{SNO}})_{\chic{\nu_e}}}{\phi^{\chic{NC}}
_{\chic{SNO}}} = 1 \; ,
\ee

\ni and from Eqs. (\ref{aa}) and (\ref{otra}), we could write
\be
\lb{3eq}
\frac{R^{\chic{CC}}_{\chic{SNO}}}{R^{\chic{ES}}_{\chic{SNO}}} \;
\frac{I_{\chic{SNO}}^{\chic{ES}}}{I_{\chic{SNO}}^{\chic{CC}}}
=\frac{R^{\chic{CC}}_{\chic{SNO}}}{R^{\chic{NC}}_{\chic{SNO}}} \;
\frac{I_{\chic{SNO}}^{\chic{NC}}}{I_{\chic{SNO}}^{\chic{CC}}} = 1 \; ,
\ee

\ni where
$
I_{\chic{SNO}}^{\chic{X}} = \overline \sigma^{\chic{X}}_{\chic{SNO}}
\langle \Pecal \rangle^{\chic{X}}_{\chic{SNO}} \; ,
$
 with $X=CC,NC,ES$.

Taking into account the equality in Eq. (\ref{3eq})
we find that the following condition should be met

\be
\lb{nueva}
\int d\En \;\varphi_{\chic{SSM}}\dEn \;\Pecal\dEn \;
\lambda_{\chic{SNO}}\dEn
= 0 \; .
\ee

\begin{figure}
\centering
\rotatebox{-90}{\scalebox{0.30}{\includegraphics{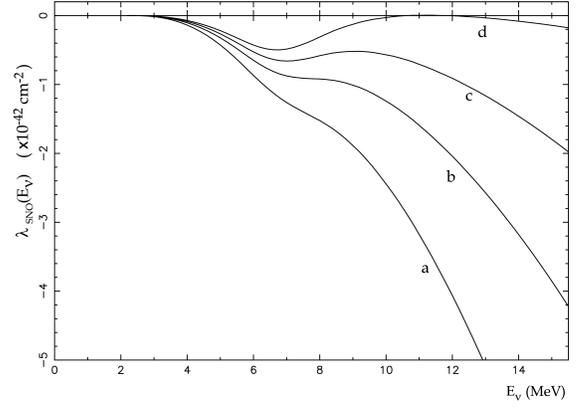}}}
\caption{{\small The function $\lambda_{\chic{SNO}}\dEn \; {\rm vs}$
the neutrino energy for
$R^{\chic{CC}}_{\chic{SNO}}/R^{\chic{NC}}_{\chic{SNO}}=4.50$ (a),
3.41 (b), 2.80 (c), 2.31 (d).}}
\label{sigmas}
\end{figure}

\ni where $\lambda_{\chic{SNO}}\dEn =
\sigccsno - \frac{R^{\chic{CC}}_{\chic{SNO}}}{R^{\chic{NC}}_{\chic{SNO}}}\;
\signcsno$. Using the values calculated by Bahcall \cite{bahcall_web} for the
CC and NC cross sections which take into account the resolution and
threshold used in SNO, it can be seen that $\lambda_{\chic{SNO}}\dEn<0 $,
for $\En>2.2 $MeV, whenever  the ratio
$\frac{R^{\chic{CC}}_{\chic{SNO}}}{R^{\chic{NC}}_{\chic{SNO}}}>2.31$.
Since $\varphi_{\chic{SSM}}\dEn \;\Pecal\dEn$ is positive then, if
the measured  ratio
$R^{\chic{CC}}_{\chic{SNO}}/R^{\chic{NC}}_{\chic{SNO}}$ is greater than
$2.31$,
the condition stated in Eq. (\ref{nueva}) cannot be
met, leading to the conclusion that $P_{a}\dEn$ cannot be equal to zero.
For reference, $E_\nu = 3.2$ MeV corresponds to an average
recoil electron kinetic energy of $5.02$ MeV, according to
\cite{bahcall_web}. Then, the integrand in Eq. (\ref{nueva}) is negative
definite in the relevant neutrino energy range if
$R^{\chic{CC}}_{\chic{SNO}}/R^{\chic{NC}}_{\chic{SNO}}>2.31$
(See Fig. \ref{sigmas}).

It is possible to estimate the unconstrained rates of SNO using the
information that has been published by the collaboration \cite{sno_nc}. The
ES unconstrained rate can be taken to be the same as that constrained by the
$^8\!$B standard shape, since it is determined essentialy from energy
independent observations (${\rm cos}\;\theta$ distribution). The NC
unconstrained rate $R^{\chic{NC}}_{\chic{SNO}}$ can be estimated in terms
of the constrained rate
$\left(R^{\chic{NC}}_{\chic{SNO}}\right)_{\chic{cons}}$ and the corresponding
total fluxes that have been reported by the collaboration:
$$
R^{\chic{NC}}_{\chic{SNO}} =
\frac{\left(\phi^{\chic{NC}}_{\chic{SNO}}\right)}
{\left(\phi^{\chic{NC}}_{\chic{SNO}}\right)_{\chic{cons}}}
\left(R^{\chic{NC}}_{\chic{SNO}}\right)_{\chic{cons}}  \; ,
$$

\ni where $\phi^{SNO}_{NC}= 6.42 \pm 1.67 \times
10^{-6}{\rm cm^{-1}s^{-1}}$ and $(\phi^{SNO}_{NC})_{\chic{cons}}= 5.09 \pm
0.63 \times 10^{-6}{\rm cm^{-1}s^{-1}}$ are the total unconstrained and
constrained NC fluxes, respectively.
Finally, the CC unconstrained rate is calculated considering that the
total number of signal events is the same as for the constrained analysis.
Taking these considerations properly into account, we estimate the ratio
of unconstrained rates to be
$$
R^{\chic{CC}}_{\chic{SNO}}/R^{\chic{NC}}_{\chic{SNO}} =
2.5 \pm 0.8  \; .
$$
 \ni The error is large because the error in the estimate of the NC
unconstrained rate in terms of the unconstrained total NC flux is large.
Nontheless, the central value is well above the lower limit of 2.31
given above, and indicates that the need for active
oscillations is favored.

If the forthcomming results from SNO confirm that
$R^{\chic{CC}}_{\chic{SNO}}/R^{\chic{NC}}_{\chic{SNO}}$ is
actually larger than the limit we found using the estimates of
\cite{bahcall_web} for the (response-averaged) cross-section, then
the probability transition of solar $\nu_e$ into an active
neutrino must be different from zero. Consequently, it is not
possible to explain the experimental CC and NC results of the
collaboration claiming only spectral distortion at the Earth
and/or oscillations into sterile neutrinos. It is important to
notice that we arrived to this conclusion without assuming that
$\Peacal\dEn$ are constant.

A systematic calculation of the shape of the $^8\!$B neutrino
spectrum has been presented in \cite{shape}, together with an
estimation of the theoretical and experimental uncertainties. No
such precise knowledge has been required in our approach,  based
in the analysis of the negativeness of the integrand in Eq.
(\ref{nueva}).

\section{Model Independent analysis of SK and SNO}
\lb{ultima}

In this section, we will use the elastic scattering measurement of
SK instead of the corresponding measurement of SNO because it has
a smaller error. Equivalently to Eq. (\ref{gg}), we have
\be
\lb{jj}
R^{\chic{ES}}_{\chic{SK}} = \overline\sigma^{\;\chic{ES}}
_{\chic{SK}}\;
\phi^{\chic{ES}}_{\chic{SK}} \; ,
\ee

\ni with definitions like those given in Eq. (\ref{prob_av_es}).

 As noted by Fogli {\it et al.} \cite{fogli}, the
response functions of SNO and SK behave quite similarly if
appropiate thresholds are used.
In this way the equality of $\langle \Peacal\rangle^{ES}_{SK}$ and
$\langle \Peacal \rangle^{CC}_{SNO}$ can be ensured.  As discussed in the
previous section and noticed in ref. \cite{bandy1}, this equality can also
be stablished independently of the kinetic energy threshold if
the energy independence of the $\Peacal\dEn$ is adopted.
Here we follow this approach. Accordingly, Eqs. (\ref{ff}), (\ref{ee}),
and (\ref{jj}) can be rewritten as follows:
\bea
\lb{las_ecs}
r_{\chic{ES}}              & =& x + \rho\; y \; ,  \nn \\
r_{\chic{CC}} & =& x      \; ,       \nn \\
r_{\chic{NC}} & =& x + y   \; ,
\eea

\ni where $r_{\chic{ES}}$, $r_{\chic{CC}}$, and
$r_{\chic{NC}}$ are the total rates normalised to the SSM
predictions:
\bea
R_{\chic{SSM}}^{\chic{ES}} &=&
\overline \sigma^{\chic{ES}}_{\chic{SK}} \;\phi_{\chic{SSM}} \; ,\nn \\
R_{\chic{SSM}}^{\chic{CC,NC}} &=&
\overline\sigma^{\chic{CC,NC}}_{\chic{SNO}} \;\phi_{\chic{SSM}} \; .
\eea

\ni We have introduced the variables
$x\equiv f\Pecal$ and $y\equiv f\Pacal$,
which represent  the relevant degrees of freedom of the problem.
Since $\Peacal$ are constants, then $0\leq \Pecal + \Pacal \leq 1$.

 From Eq. (\ref{las_ecs}), $r_{\chic{NC}}$ can be
expressed in terms
of $r_{\chic{CC}}$ and $r_{\chic{ES}}$ :
\be
\lb{univ}
r_{\chic{NC}} =
\frac{1}{\rho}\left[ r_{\chic{ES}} -(1-\rho)
\;r_{\chic{CC}}
\right]  \; ,
\ee

\ni which is valid for any value of $x$ and $y$ \cite{barger}.

 We define the $\chi^2$ function
\be
\lb{chi}
\chi^2 = \sum_{X} \frac{(r_X\med{(x,y)}- r^{\chic{exp}}_X)^2}
{\sigma_X^2} \; ,
\ee

\ni where $r_X\med{(x,y)}$ are given in  Eq. (\ref{las_ecs}). Here,
$r^{\chic{exp}}_X$ and
$\sigma_X$ are the experimental values for the normalised rates and
their errors respectively \cite{sno_nc,sno_cc,superk}:
\bea
\lb{valores}
r^{\chic{exp}}_{\chic{SK}}  &=& 0.459 \pm 0.017 \nn \\
r^{\chic{exp}}_{\chic{CC}}  &=& 0.349 \pm 0.021 \nn \\
r^{\chic{exp}}_{\chic{NC}}  &=& 1.008 \pm 0.123  \; .
\eea


\begin{figure}
\centering
\rotatebox{-90}{\scalebox{0.30}{\includegraphics{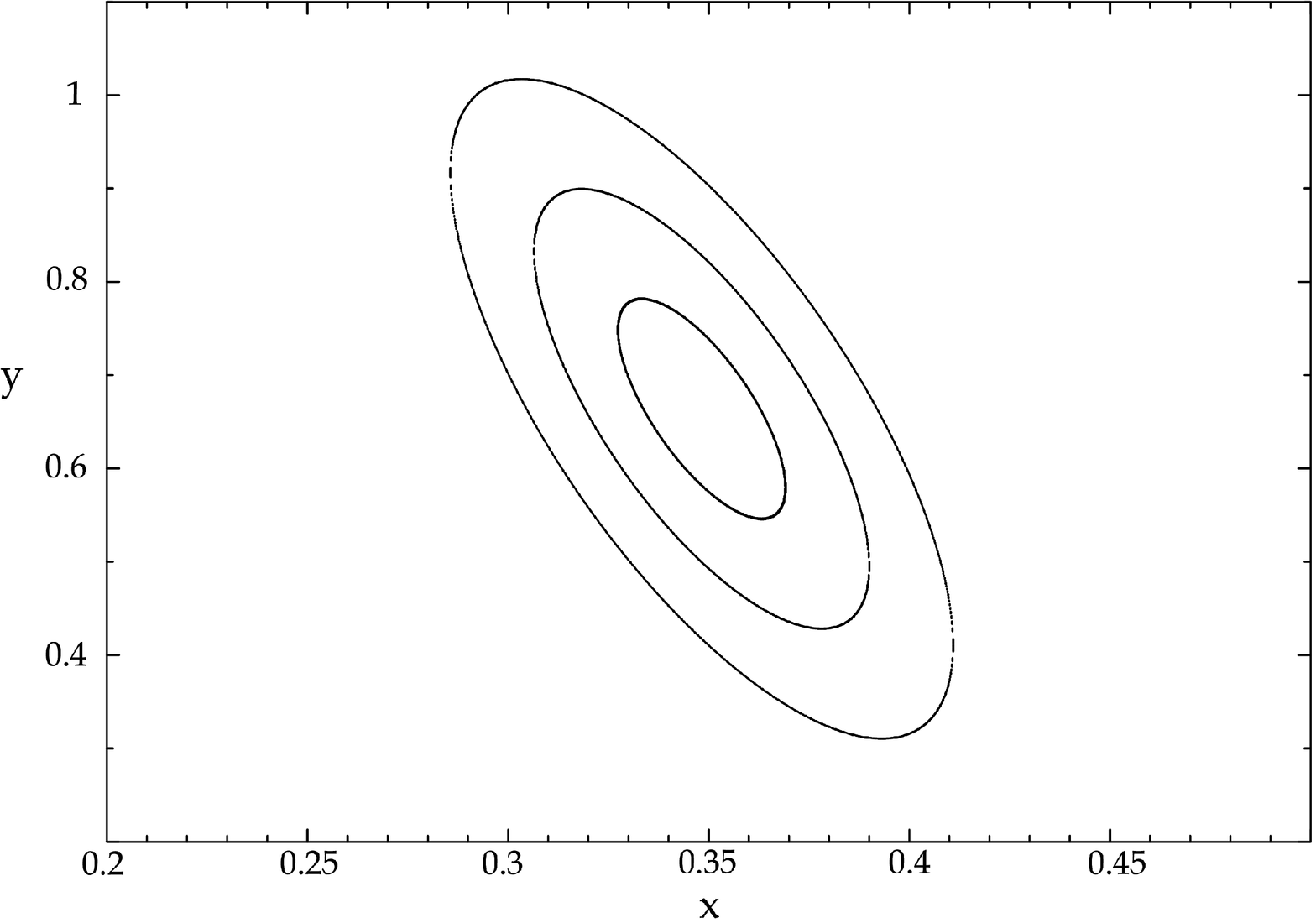}}}
\caption{{\small Contours for $\Delta \chi^2 = 1,\;4$, and 9 of
allowed values for $x=f\;\Pecal$ and $y=f\;\Pacal$.}}
\label{xy}
\end{figure}

\begin{figure}
\centering
\rotatebox{-90}{\scalebox{0.30}{\includegraphics{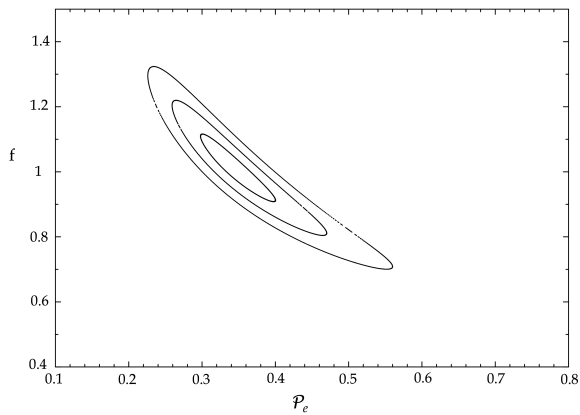}}}
\caption{{\small Contours for $\Delta \chi^2 = 1,\;4$, and 9 of
allowed values for $f$ and $\Pecal$ obtained from mapping the
corresponding contours of Fig. \ref{xy} with the condition
$\Pecal+\Pacal=1$.}}
\label{fpe}
\end{figure}

\begin{figure}
\centering
\rotatebox{-90}{\scalebox{0.30}{\includegraphics{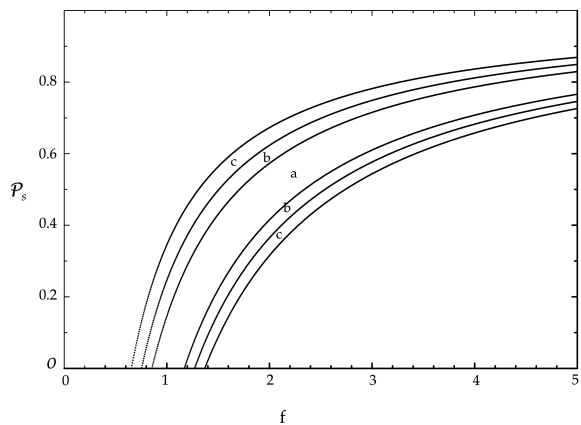}}}
\caption{{\small Regions of allowed values of $f$ and $\Pscal$ at
(a) 68$\%$; (b) 95$\%$, and (c) 99$\%$ C.L.}}
\label{fps}
\end{figure}

Letting  $x$ and $y$ vary as free parameters, we find the minimum
value of $\chi^2$ ($\chi^2_{min}= 0.0039$), and the
$\Delta \chi^2 =1,\;4$, and 9 contours for these parameters as shown
in Fig. \ref{xy}. The projection of these contours on the $x$, and $y$
axes ($N_{DF}=1$ in each case), give their $1\sigma$, $2\sigma$, and
$3\sigma$ ranges \cite{rpp}. The best fit values along with their
$1\sigma$ errors are
\bea
x &=& 0.35\pm 0.02 \; ,\nn \\
y &=& 0.66 \pm 0.11 \; .
\lb{x_y_best_fit}
\eea

\ni The previous values times the SSM total $^8\!$B flux
($\phi_{\chic{SSM}}= 5.05 \times 10^6 {\rm cm}^{-2}s^{-1}$), give the
$\nu_e$ and $\nu_a$
components of the flux which are consistent with the values
reported by SNO \cite{sno_nc}.

When $f=1$, {\it i.e.}, there is no discrepancy between the
SSM and the true total $^8\!$B neutrino flux, Ec.(\ref{x_y_best_fit}) gives
the 1$\sigma$ ranges for the quantities $\Pecal$ and $\Pacal$. In this case
the sum $\Pecal+\Pacal$ is consistent with being equal to one.

Let us now assume that there exist oscillations only among active
states. Then, we have $\Pecal + \Pacal =1$, there is also no deformation
of the spectrum produced in the Sun ($\zeta\dEn = 1$), and  $f=x+y$.
We obtain the
$1\sigma$, $2\sigma$, and $3\sigma$ ranges for $f$ and $\Pecal$
from the contours in Fig. \ref{fpe}, built by
mapping the contours of Fig. \ref{xy} to the plane $f\;{\rm vs}\;\Pecal$
using the constriction $\Pecal+\Pacal=1$. These contours coincide with those
found in ref \cite{bandy1} directly from Eq. (\ref{las_ecs}), with $y$
replaced by $f-x$.

From Fig. \ref{fpe} it can be seen that, by including the
NC measurement in the analysis, a significant improovement
has been achieved in the $1\sigma$ error bar of $f$ with respect to
the one obtained using only the SK and the SNO CC data \cite{fogli}.
The best fit values and their $1\sigma$ ranges are
\bea
f&=&1.01^{+0.11}_{-0.09}  \nn \\
\Pecal&=&0.34^{+0.05}_{-0.04}
\eea

Impossing the less stringent condition
$\alpha\leq (\Pecal+\Pacal)\leq 1$,
with $0\leq \alpha \leq 1$ the value of $f$ will be bounded by
\be
x+y\leq f \leq \frac{x+y}{\alpha} \; ,
\ee

\ni from where we see that allowing for a non vanishing probability to
oscillate into a sterile neutrino ($\alpha \neq 1$), we have
larger upper bound for $f$.
Assuming that $\Pecal+\Pacal+\Pscal=1$, we have that
\be
\Pscal = 1 -\frac{x+y}{f} \; .
\ee

\ni From the dispersion of $x$ and $y$ we can find allowed
regions in the $\Pscal \; {\rm vs} \; f$ plane, corresponding to the 68,
95, and 99 $\%$ confidence levels. As shown in Fig. \ref{fps},
these regions are not bounded and hence it is not possible to
determine $f$ and $\Pscal$ with the existing data \cite{barger}.

\section{Conclusions}

In this work we have examined the relation between the observed
quantities $\phi^{\chic{CC}}_{\chic{SNO}}/\phi^{\chic{NC}}_{\chic{SNO}}$,
$\phi^{\chic{CC}}_{\chic{SNO}}/\phi^{\chic{ES}}_{\chic{SNO}}$ with the
flavor fractional $\nu_e$ content of the fluxes measured through the
ES and NC reactions. When combined with the hypothesis of a non
distorted $^8\!$B spectrum the measurement gives a clear signal
of active flavor transformation. As we also show, when available, the
SNO experimental rates unconstrained by the $^8\!$B standard shape,
combined with the cross-section as calculated in ref.\cite{bahcall_web},
could give conclusive evidence for active oscillations, even for a non
constant $\Pecal\dEn$.

Finally a model independent analysis including the latest SK and SNO data
is performed under the assumption of  constant $\Peacal\dEn$, with and
without the condition $\Pecal + \Pacal =1$. Our result agrees with
ref. \cite{barger} in the sense that no conclusion can be drawn with the
present data about the sterile neutrino content of the solar neutrino flux.


\begin{acknowledgments}
This work has been partially supported by PAPIIT-UNAM Grant
IN109001 and by CONACYT Grants 32279E and 35792E . The authors
wish to thank R. Van de Water for useful comments.
\end{acknowledgments}


\begin{thebibliography}{200}

\bibitem{sno_nc}{The SNO Collaboration, R. Ahmad {\it et al.}
Phys. Rev. Lett. {\bf 7}, 649 (2002).}

\bibitem{ortiz}{C. E. Ortiz, A. Garcia, R. A. Waltz, M. Bhattacharya,
and A. K. Komives, Phys. Rev. Lett. {\bf 85}, 2909 (2000).}

\bibitem{sno_cc}{The SNO Collaboration, R. Ahmad {\it et al.}
Phys. Rev. Lett.  {\bf 87}, 071301 (2001).}

\bibitem{superk}{Super-Kamiokande Collaboration, S. Fukuda
{\it et al.}, Phys. Rev. Lett. {\bf 86}, 5651 (2001).}

\bibitem{global}{J. N. Bahcall, M. C. Gonzalez-Garcia, and C.
Pe\~na-Garay, J. High Energy Phys. {\bf 07}, 054 (2002); P. C.
Holanda and A. Yu. Smirnov hep-ph/0205241.}

\bibitem{bandy1}{A. Bandyopadhyay, S. Choubey, S. Goswami, and,
D.P. Roy, Phys. Lett. B {\bf 540}, 14 (2002).}

\bibitem{bp:2001}{J. N. Bahcall, S. Basu, and M. H. Pissoneault,
Astrophys. J. {\bf 55}, 990 (2001).}

\bibitem{bahcall_web}{We used the tabulated values for the
cross sections found at the URL http://www.sns.ias.edu/ ~ jnb/SNdata.}

\bibitem{fogli}{G. L. Fogli, E. Lisi, D. Montanino, and, A. Palazzo,
Phys. Rev. D {\bf 64}, 093007 (2001), {\it ibid.}, {\bf 65},
117301, (2002).}

\bibitem{rpp}{Particle Data Group, D. E. Groom {\it et al.},
Eur. Phys. J. High Energy Phys. {\bf 05}, 015 (2001);
Probability and Statistics sections.}

\bibitem{barger}{V. Barger, D. Marfatia, and K. Whinsant,
 Phys. Rev. Lett. {\bf 88}, 011302 (2002).}

\bibitem{shape}{J. N. Bahcall, E. Lisi, D. E. Alburger,
L. De Braeckleer, S. J. Freedman, and J. Napolitano, Phys. Rev. C
{\bf 54}, 411 (1996).}

\bibitem{n1}{P. Creminelli, G. Signorelli, A. Sturmia,
, hep-ph/0102234, v3 22 April 2002 (addendum 2).}

\bibitem{n2}{P. Aliani et al, hep-ph/0205053}

\bibitem{n3}{A. Sturmia, C. Cattadori, N. Ferrari, 
F.Vissani, hep-ph/0205261.}

\bibitem{n4}{G.L. Fogli, E. Lisi, A. Marrone, D. Montanino,
 A. Palazzo, Phys. Rev. D 66, 053010 (2002).}

\bibitem{n5}{M. Maltoni, T. Schwetz, M.A. Tortola, 
J.W.F Valle, hep-ph/0207227.}


\end{thebibliography}




\end{document}